\documentclass[aps,english,prb,reprint,superscriptaddress, 
citeautoscript,showpacs]{revtex4-1}
\usepackage{graphicx}
\usepackage{amssymb}
\usepackage{babel}
\usepackage{color}
\usepackage{multirow}
\usepackage{setspace}
\usepackage{longtable}

\begin{document}

\title{First-Principles Prediction of Graphene-Like XBi (X=Si, Ge, Sn) Nanosheets}

\author{A. Bafekry}\email{bafekry.asad@gmail.com}
\affiliation{Department of Physics, University of Guilan, 41335-1914 Rasht, Iran}
\author{M. Yagmurcukardes}
\affiliation{Department of Physics, University of Antwerp, Groenenborgerlaan 171, B-2020 Antwerp, Belgium}
\author{B. Akgenc}
\affiliation{Department of Physics, Kirklareli University, Kirklareli, 39100, Turkey}
\author{M. Ghergherehchi}
\affiliation{College of Electronic and Electrical Engineering, Sungkyun kwan University, Suwon, Korea}
\author{B. Mortazavi}
\affiliation{ Department of Mathematics and Physics, Leibniz Universität Hannover, Appelstraße 11, 30167 Hannover, Germany}

\begin{abstract} 

Research progress on single-layer group III monochalcogenides have been 
increasing rapidly owing to their interesting physics. Herein, we predict the dynamically stable single-layer forms of XBi 
(X=Ge, Si, or Sn) by using density functional theory calculations. Phonon band 
dispersion calculations and \textit{ab-initio} molecular dynamics simulations reveal the 
dynamical and thermal stability of predicted nanosheets. Raman spectra 
calculations indicate the existence of 5 Raman active phonon modes 3 of which 
are prominent and can be observed in a possible Raman measurement. Electronic 
band structures of the XBi single-layers investigated with and without spin-orbit coupling effects (SOC). 
Our results show that XBi single-layers show semiconducting property with the narrow band gap values without SOC. 
However only the single-layer SiBi is an indirect band gap semiconductor while GeBi and SnBi exhibit metallic behaviors by 
adding spin-orbit coupling effects. In addition, the 
calculated linear-elastic parameters indicate the soft nature of predicted monolayers. 
Moreover, our predictions for the thermoelectric properties of single-layer XBi 
reveal that SiBi is a good thermoelectric material with increasing temperature. 
Overall, it is proposed that single-layer XBi structures can be 
alternative, stable 2D single-layers with their varying electronic and thermoelectric properties.

\end{abstract}

\maketitle

\begin{figure*}[!hbt]
	\includegraphics[scale=1]{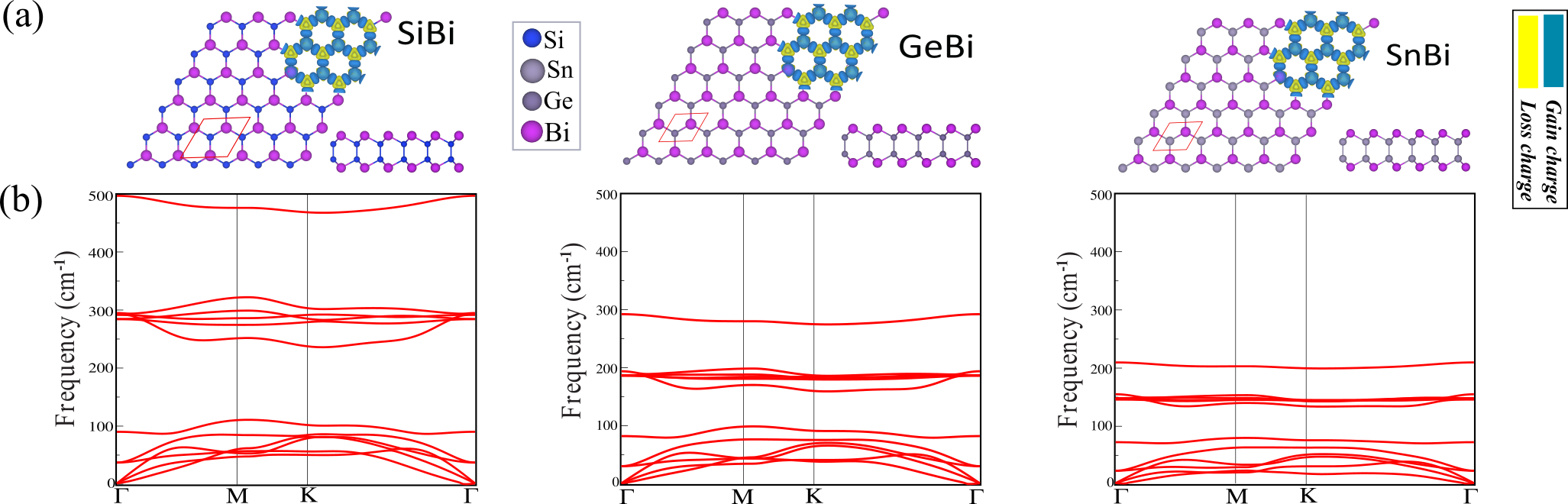}
	\caption{(Color online) For the single-layers of XBi, (a) atomic structure and (b) phonon band dispersions. 
	The primitive unit cell indicated by a red hexagonal and the difference charge density is shown in the same panel. 
	The blue and yellow regions represent the charge accumulation and depletion, respectively.}
	\label{1}
\end{figure*}
\section{Introduction}

The tremendous interests toward two dimensional (2D) materials was initiated by the graphene\cite{novoselov2004electric, geim2010rise,lee2008measurement} exceptional physics. Other famous 2D materials include Xenes (X=Si, Ge, 
Sn, P, B, and so on) \cite{cahangirov2009two, feng2017dirac, zhu2015epitaxial, 
mannix2015synthesis} and transition metal dichalcogenides (TMDs) 
\cite{xu2014spin}, has shown outstanding physical feautures to design advanced nanodevices \cite{chhowalla2013chemistry, 
tongay2012thermally}. Particularity, the electronic bands of graphene yield
Dirac cones in the vicinity of Fermi level, resulting in high carrier mobilities 
\cite{banszerus2015ultrahigh}, superior mechanical and thermal conduction properties 
\cite{papageorgiou2017mechanical}, wide band absorption \cite{liu2011plasmon}, 
and other exceptional physical/chemical features. However, semi-metallic
form of graphene has limited functionality in the semiconductor technology 
\cite{lu2013semiconducting}. In view of this, the effort of searching stable 
free-standing atomic layers of semiconducting materials continuing to look for a 
new category of 2D material class 
\cite{liu2014phosphorene,vogt2012silicene,davila2014germanene, liu2011low}. 

Very recently, bismuth-based 2D materials have attracted remarkable interest due to their unique properties.\cite{liu2020advances,xu2019two}
Bi shows very interesting features highly appealing for energy related 
applications. In fact, the electronic features of Bi-based
nanomembranes can be easily modified via introduction of distinct 
anions and cations into the layered
structure and the band gap can be tuned from 0.3 eV (near infrared absorption range) to 3.6 eV (ultraviolet absorption 
range) 
\cite{li2017activation,he2018review}. Moreover, Bi-based nanosheets belong to 
the anisotropic \textit{p} and \textit{s-p} hybridization which induces remarkably dispersed 
electronic structure \cite{xu2018sp}. Feng et al. summarized reports on the different 
strategies for using as the highly efficient visible-light photocatalysts by 
modifying electronic band structure with anisotropic \textit{p} and \textit{s-p} hybridization 
states \cite{feng2017efficient}. They showed that highly dispersed electronic 
structure of Bi-layered components not only
decreases the effective mass of photoexcited charge carriers and consequently enhance  
mobility, but also improve the charge separation and transmission efficiency in 
photoexcitation process, highly desirable for the employment in solar cells, thermoelectric and 
optoelectronic energy conversion devices \cite{schlitz2014solubility, 
zhang2020exploring}.
Thirdly, Bi-layered materials allow foreign ions to intercalate and to form 
multicomponent stable compounds without significant structural
deformation due to their stable skeleton structure
and a large interlayer space \cite{zhang2016semiconducting}.

The another family members of 2D materials are binary compounds of a group IV 
element (Si,Ge,Sn) and group-III monochalcogenides with a representative chemical formula of MX (M=B, Ga, Al, 
In and X=O, S, Se, Te) \cite{xu2017electronic, 
barreteau2016high,demirci2017structural,lin2017single,
mogulkoc2019characterization,eren2019vertical} which have honeycomb lattice and  
that effectively consists of covalently bonded atomic planes are held together 
by van der Waals interactions as well as in TMDs have been extensively studied 
for decades due to their outstanding properties, such as remarkably high carrier mobility,  p-type electronic response,
sombrero-shape valence band edges, etc. To 
date, various MX systems such as InSe, GaS, GaSe, and GaTe have been 
experimentally realized and theoretical studies have reported a stable class of
single-layer group IV monochalcogenides (YX, Y = Si, Ge,
Sn and X = S, Se, and Te) that are semiconductors with wide band gaps.
Due to above reason, we have selected silicon (Si), germanium (Ge), and tin (Sn) 
as chalcogen. It should be noted that anode materials based on the 
on Si, Ge, Sn show outstandingly high specific storage capacities of 4200, 1625, and 994 mAh$g^{-1}$ \textcolor{blue}, respectively, very appealing for the design of more efficient rechargeable batteries \cite{li2017si}.

Motivated with the recent realization of 2D MX single-layers and their novel 
properties, se here study the vibrational, mechanical, electronic and thermal transport properties of XBi (X= Si, Ge, and Sn) 
single-layers by utilizing first principles for the first time. Our results improve the understanding  
on the importance of the chemical composition and structural configuration of XBi (X= Si, Ge, and Sn) nanosheets 
and may hopefully guide experimental studies.

\section{Method}
 We conducted density-functional theory (DFT) simulations using the projector augmented wave (PAW) and generalized
 gradient approximation (GGA) proposed by the Perdew-Burke-Ernzerhof (PBE) form\cite{GGA-PBE1,GGA-PBE2}
 employing Vienna \textit{ab-initio} Simulation Package (VASP)\cite{vasp1,vasp2}. The vdW dispersion correction was applied via
 the DFT-D2 method of Grimme\cite{Grimme}. Spin-orbit-coupling (SOC) was taken into account for
 the electronic-band structure calculations. A kinetic energy cut-off of 600 eV was considered in DFT calculations for the plane-waves \cite{kresse1999ultrasoft}.
 The hybrid functionals (HSE06) \cite{paier2006screened} with (SOC) were also taken into account as called HSE+SOC.
 The Brilliouin zone sampled by using a $\Gamma$-centered 16$\times$16$\times$1 \textit{k}-point mesh
 for the unit cell. The geometries were optimized until the energy difference
 between two following steps were less than 10$^{-5}$ eV, and maximum force on atoms was smaller than 10$^{-3}$ eV \AA{}$^{-1}$. A $\sim$20 \AA{} vacuum was also applied along the sheet's normal direction to avoid the inaccuracies due to the interactions with monolayer images.
 The charge transfers are evaluated by the decomposition of charge density on the basis of Bader charge method \cite{henkelman2006fast}. The vibrational properties and the phonon dispersion relations were acquired via the small-displacement method using the PHONOPY code\cite{alfe2009phon}. \textit{Ab-initio} molecular dynamics (AIMD) simulations were also carried out to examine the thermal stability of XBi single-layers by using 4$\times$4$\times$1 super cell at room temperature (300 K) with total simulation time of 6 ps with 2 fs time steps. 

\section{Results and Discussions}

\subsection{Structure and Stability}
The atomic structure of XBi single-layers consists of four three-coordinated X (Si, Ge and Sn) and four-fold coordinated Bi atoms in a hexagonal unit cell 
containing four atoms, as shown in Fig. \ref{1}(a). 
In the a single-layer structure 
2-X layers are sandwiched between Bi-layers in the Bi-X-X-Bi order. 
The optimized lattice constants, $a=b$, are calculated to be 4.09, 4.15, and 4.35 \AA {}, for SiBi, GeBi, and SnBi, respectively, 
which are slightly larger than those of 
Ga-monochalcogenides (3.58 and 3.75 \AA {} for single-layers of GaS and GaSe 
structures, respectively).
The bond lengths of $d_{1}$ (X-Bi) are found to be 2.69 \AA{} (Si-Bi), 2.74 \AA{} 
(Ge-Bi) and 2.90 \AA{} (Sn-Bi), while the $d_{2}$ (X-X) are slightly smaller (2.31 \AA{} (Si-Si), 2.43 \AA{} (Ge-Ge) and 2.80 \AA{} (Sn-Sn)). 
The charge density difference of XBi (X=Si, Ge, Sn) single-layers is shown in Fig. 
\ref{1}(a) in the same panel, in which yellow and blue color coding represent the 
charge depletion and accumulation, respectively.  
It is clear that Bi atoms are charged negatively and surrounded by X (Si, Ge and Sn) atoms that are positively charged, which reveal the charge transfer from X atoms to the connecting Biatom.  
The difference charge density ($\Delta \rho$) is defined as:
\begin{equation}
\Delta\rho =\rho_{XBi}-\rho_{X}-\rho_{Bi}
\end{equation} 
where $\rho_{XBi}$, $\rho_{X}$ and $\rho_{Bi}$ represents the charge densities 
of the XBi and isolated atoms, respectively. 
Notice that each Bi atom gains about 0.04, 0.17 and 0.20 $e$ from the adjacent 
Si, Ge and Sn atom in the SiBi, GeBi and SnBi single-layers, respectively. The 
charge redistribution is due to the different electro-negativities of 1.9 (Si), 
2 (Ge), 1.96 (Sn) and 2.02 (Bi).
The structural and electronic parameters of the XBi (X=Si, Ge, Sn) single-layers 
are listed in Table I.
In order to calculate the cohesive energy $E_{coh}$, the used expression as follows:
\begin{equation}
E_{coh} = \frac{2E_{X}-2E_{Bi}-E_{tot}}{n_{tot}}
\end{equation}
where $E_{X}$ and $E_{Bi}$ represent the energies of isolated single X (Si, Ge 
and Sn) and Bi atoms, n$_{tot}$ is the total number of unit cell, respectively; $E_{tot}$ represents the total energy of the 
XBi single-layer.
The cohesive energy of SiBi, GeBi and SnBi, are found to be 4.65, 4.32 and 4.06 
eV/atom, respectively. As a matter of fact, the more negative values for cohesive energies confirm the more stability and thus the stability is higher for the lattices with a lighter X atom. 
We further analyze the formation 
energy of the predicted structures using the formula below;

\begin{equation}
E_{for} = E_{bulk}(XBi)/layer-E_{bulk}^X-E_{bulk}^Bi
\end{equation} where $E_{bulk}(XBi)$, $E_{Bulk}^X$, and $E_{Bulk}^Bi$ represent the 
total energies of bulk form of XBi structure and that of the individual atoms. The calculated formation energies are 0.35, 0.54, and 0.59 eV for GeBi, SiBi, and 
SnBi single-layers, respectively. The positive formation energies indicate the chemical instability of the predicted structures as if they are formed from 
the bulk XBi structures. In contrast, these results suggest the formation single-layer XBi from the isolated X and Bi atoms.

\begin{figure}[!t]
	\includegraphics[scale=1]{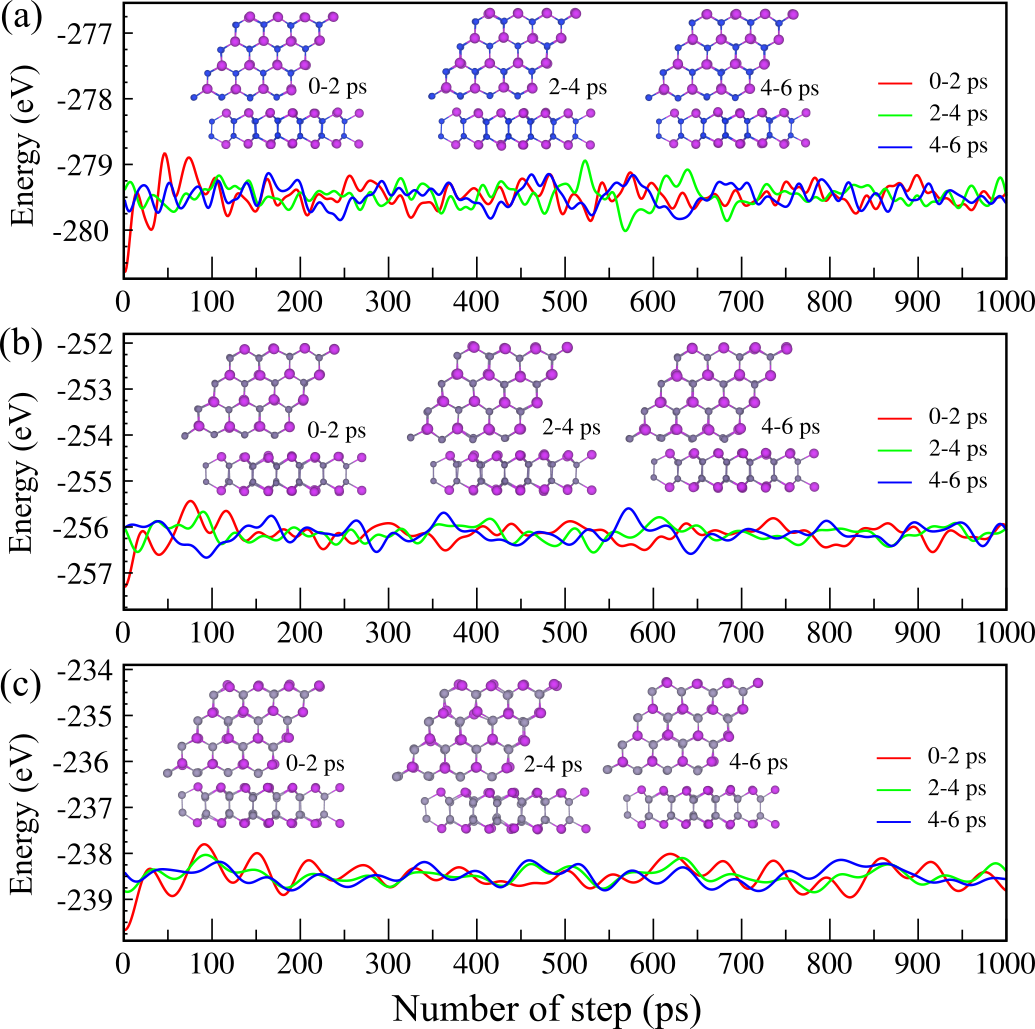}
	\caption{(Color online) AIMD simulation of (a) SiBi, (b) GeBi and (c) SnBi single-layers. The snap shut of optimized structures are indicated in the inset.}
	\label{2}
\end{figure}

\begin{figure}[!b]
	\includegraphics[scale=1]{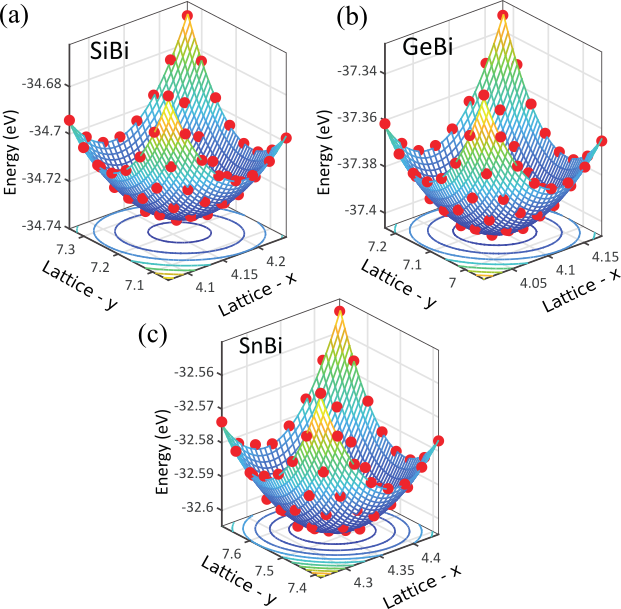}
	\caption{Energy landscapes for the (a) SiBi, (b) GeBi and (c) SnBi single-layers.}
	\label{3}
\end{figure}

\begin{table*}[!htb]
	\centering
	\caption{\label{table1} The calculated values for the optimized XBi single-layers: lattice constants \textit{a}, bond distances (the bond length between X-Bi atoms $d_{X-Bi}$ and X-X atoms $d_{X-X}$, where X=Si, Ge 
		and Sn), bond angles between Bi-X-Bi atoms ($\theta_{1}$) and Bi-X-X ($\theta_{2}$), cohesive energy, charge differences (according to Bader analysis), electronic states $(ES)$ are specified as metal (M) semiconductor (SC) and band gap energy (PBE / PBE+SOC / HSE+SOC)}
	\begin{tabular}{lcccccccccccc} 
		\hline\hline
		&\textit{a} & \textit{d$_{X-Bi}$}& \textit{d$_{X-X}$}& 
		\textit{$\Delta{z}$}& \textit{$\theta_{1}$}& \textit{$\theta_{2}$}& $E_{coh}$ & $\Delta{Q}$ & $ES$ & $E_{g}$ \\ 
		& (\AA) & (\AA) & (\AA) & (\AA) & ($^{\circ}$) & ($^{\circ}$) & (eV/atom) & ($e$) & (eV) & (eV)\\
		\hline
		SiBi  & 4.09  & 2.69  & 2.31  & 4.89 &  98.94  & 118.63  & 4.65 & 0.04 & SC & 
		(0.71/0.25/0.65) \\
		GeBi  & 4.15  & 2.74  & 2.43  & 5.08 &  98.65  & 118.86  & 4.32 & 0.17 & M & (0.38/0/0) 
		\\
		SnBi  & 4.35  & 2.90  & 2.80  & 5.71 &  97.03  & 120.11  & 4.06 & 0.20 & M & (0.30/0/0) 
		\\
		\hline\hline
	\end{tabular}
\end{table*}

\begin{figure*}[!htb]
	\includegraphics[scale=1]{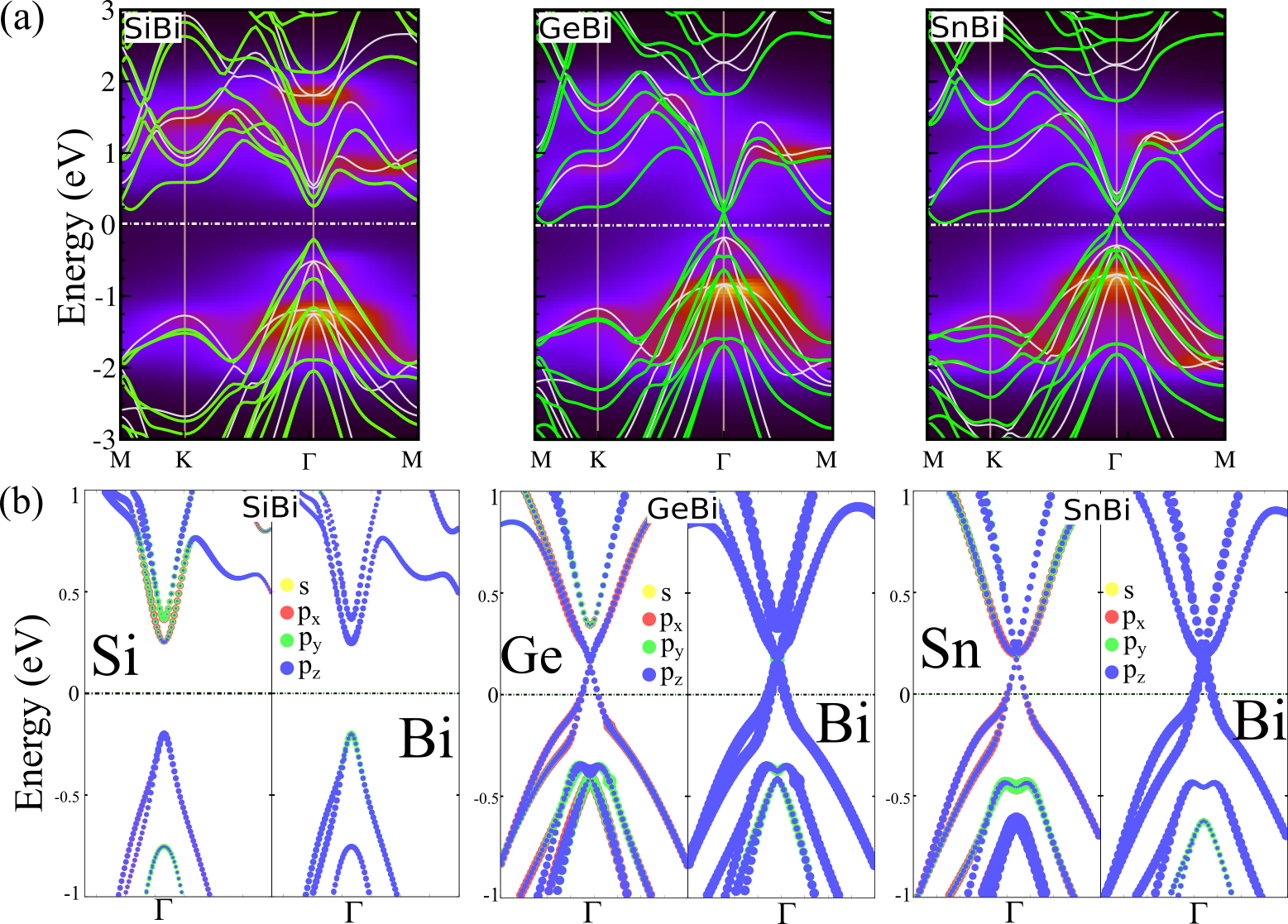}
	\caption{ (a) Intensity map of XBi (X=Si, Ge, Sn) single-layers with corresponding electronic band structure which overlayed by green line and (b) 
		orbital resolved band structure with HSE+SOC of XBi single-layers.}
	\label{4}
\end{figure*}

The dynamical stability of XBi monolayers is investigated by calculating their 
phonon band dispersions through the whole BZ which are presented in Figs. 
\ref{1}(b). Apparently, phonon branches are free from any imaginary frequencies 
indicating the dynamical stability of the structures. Similar to the case of 
single-layer InSe, which has the same symmetry and crystal structure with those 
of XBi, XBi single-layers exhibit three acoustic and nine optical phonon 
branches. 
Among nine of the optical branches, three of them are found to be non-degenerate 
out-of-plane vibrational modes while the remaining six are three different 
doubly-degenerate phonon modes. The calculated Raman spectra reveal that each 
single-layer structure exhibits three prominent Raman active modes which are 
described as follows: The highest frequency Raman active phonon mode has 
out-of-plane vibrational character. This mode is assigned to the opposite 
out-of-plane vibration of X-layers 
and Bi-layers, respectively. The frequencies of the mode are calculated to be 
207, 291, and 496 cm$^{-1}$ for single-layers SnBi, GeBi, and SiBi, 
respectively. 
In addition, the most prominent Raman active phonon mode is found to exhibit 
in-plane vibrational behavior. In this doubly-degenerate phonon mode, the X and 
Bi atoms vibrate out-of-phase in the in-plane directions that can be 
interpretted as the shear-like motion of X and Bi layers. The frequencies are 
found to be 146, 186, and 283 cm$^{-1}$ for SnBi, GeBi, and SiBi, respectively. 
Moreover, each structure exhibits also an out-of-plane Raman active phonon mode 
which has relatively smaller intensity and frequency. The mode is assigned to 
the out-of-plane breathing-like vibration of top and bottom X-Bi sublayers. Its 
frequency is found to be 71, 82, and 89 cm$^{-1}$ for SnBi, GeBi, and SiBi, 
respectively.

In addition, the thermal stability are examined by performing \textit{ab-initio} 
molecular dynamics (AIMD) simulations using NVT ensemble with fixed particle number, volume and temperature. 
For the AIMD simulations, 32-atom supercell is used for each single-layer with a 
$k$-mesh of 4$\times$4$\times$1. The dynamical investigations are started with the optimized structure of XBi single-layers at 0 K and discussed in Structure ans stabiliy of XBi single-layers section. We have further extened our calculations to the thermal stabiliy at room temperature. During the simulations, temperature is 
kept at 300 K, the fluctuations of total energy and evolutions of XBi atomic structures 
during the simulations are shown in Figs. \ref{2}(a-c). The time step was set to 2 fs and, to reach a total simulation time of 6 ps, 
1000 steps were realized three times. Due to the large size of the cell, all the calculations were performed with each 2 ps. 
The structure snapshots are taken at the end of the each simulation in every 2 ps.
As can be seen in Fig. \ref{2}, none of the single-layers undergo structural reconstruction even around 300 K indicating the 
thermal stability of each single-layer.
The variation of total energy per atom is 1 eV which is in the acceptable range similar to many studies in the literature. 
In addition, as the X atoms changes from Si to Sn, the single-layer structure displays in-plane buckling with 
increasing temperature due to different X-Bi and Bi-Bi bond formations. It can be concluded that XBi single-layers exhibit 
thermal stability around the room temperature.

Another analysis to test the stability of the predicted material is the 
investigation of the mechanical properties. For this purpose, we firstly change the 
unitcells of the XBi structures to the rectangular cell and applied sequential 
uniaxial strains. The applied maximum strain ratio is $\mp$2\% of the relavent 
lattice parameter of the rectangular cells. Obtained energies versus lattice 
parameters (energy landscape) are given in Figs.~\ref{3}(a-c). To calculate the 
in-plane stiffness\cite{topsakal} we used $C_{x,y}$=$(1/A)\partial 
E^{2}_{T}/\partial\epsilon^{2}_{x,y}$ and to find Poisson's ratio we used 
$\nu_{x,y}=\epsilon_{y}/\epsilon_{x}$ equations. In here, $A$ defined the 
unitcell area, $E_{T}$ is for the total energy per cell of the XBi structure, 
and the applied uniaxial strain along the $x,y$ axis is defined as 
$\epsilon_{x,y}$. Because of equivalent in-plane lattice constants of the XBi 
structures, we found that $C_x = C_y$ and $\nu_{xy} = \nu_{yx}$. The calculated 
in-plane stiffness values are 76.96, 64.13 and 45.72 J/m$^2$ for SiBi, GeBi and 
SnBi, respectively while the obtained Poisson's ratio values are 0.279, 0.295, 
and 0.290. These results indicate that in-plane stiffness of the XBi structures 
get smaller while going from SiBi to SnBi. However, these in-plane stiffness 
values are larger than their individual components such as Silicene, Germanene, 
Stanene and Bismuthene,\cite{ethem1,ethembi,fatihso} while these numbers are 
almost equal to that of many transition metal dichalcogenides.\cite{mehmetPtX2,fatihRuX2} 

\subsection{Electronic Properties}

The intensity map band structure of XBi single-layers is shown in Fig. \ref{4}(a) 
which overlayed with its electronic band structure with considering spin orbital 
coupling (SOC). XBi single-layers exhibit semiconducting behavior without 
SOC effects and the calculated band gaps are  0.71, 0.38 and 0.30 eV for SiBi, GeBi and SnBi, respectively. 
However, inclusion of SOC shows that only the SiBi single-layer is a semiconductor with an indirect band gap of 0.25 eV under 
PBE+SOC functional. In addition, the conduction band minimum (CBM) of SiBi resides between the $M-K$ points while the valence band maximum (VBM) 
lies at the $\Gamma$-point. The HSE06 functional is also used to evaluate the energy band gap. We see that 
the HSE06 approach does not change the sort of indirect band gap however, it gives rise to an increase of the band gap to 0.65 eV. 
The calculated electron effective mass along $\Gamma$ 
$\longrightarrow$ K (M) is 0.15 (0.29) $m_{e}$$^*$, whereas the hole effective 
mass is estimated to be -0.1 and -0.16 $m_{e}$$^*$ along $\Gamma$ $\longrightarrow$ K and 
$\Gamma$ $\longrightarrow$ M, respectively. These estimated light effective masses for the electron and hole 
 confirm high carriers mobility in these novel nanosheets. 
In contrast SiBi single-layer, we find that the GeBi and SnBi single-layers show a metallic 
characteristic. 
To better examine the contribution of different orbitals to the electronic 
states and the bonding characteristics of XBi single-layers, we carry out the 
calculations of the orbital-resolved band structure as shown in Fig. \ref{4}(b)
It is conspicuous that the states near the Fermi energy shows contributions from $p$ 
orbitals of X and Bi atoms. It is clear that X and Bi atoms $p_{z}$ orbitals contributions are much higher than that from $p_{x,y}$-orbitals. The fact that the 
$p_{z}$-orbitals are dominant is caused by the $sp^{3}$-like bond of X and the 
$sp^{2}$-like bond of Bi forming the SiBi single-layer. In addition, the contribution of $p_{z}$-orbitals can also allow the surface oxidation of 
the predicted structures. 
From Fig. \ref{2}(b) is clearly seen that the metallicity of GeBi and SnBi 
single-layers mainly are composed of the $p_{z}$ orbitals of Bi atom, while the 
contribution of Ge or Sn atom orbitals do not contribute. 

\subsection{Thermoelectric Properties}
\begin{figure*}[!htb]
	\includegraphics[scale=1]{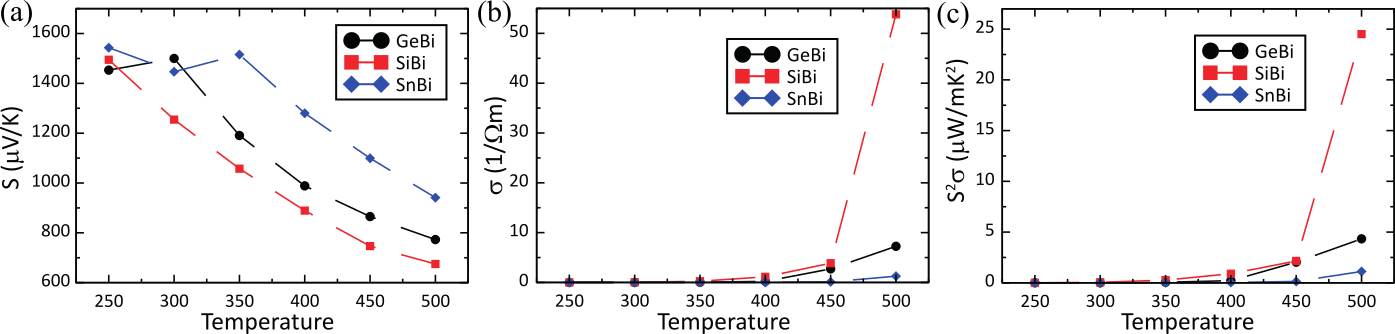}
	\caption{(Color online) The calculated Seebeck coefficient ($S$), electrical conductivity $\sigma$ and power factor ($S^2 \sigma$) of XBi single-layers for various temperatures from 250K to 500K.}	
	\label{5}
\end{figure*}

In order to estimate the thermoelectric properties of the XBi single-layers, the 
semiclassical Boltzmann transport theory (BTT) within the constant relaxation 
time approximation and the rigid band approach as implemented in the BoltzTraP2 
code was used. \cite{MADSEN2018140,MADSEN200667}
By using this package and based on the electronic structure, the 
Seebeck coefficient ($S$), electrical conductivity ($\sigma$) and power factor 
($S^2 \sigma$) are all calculated. Obtained results are given with respect to relaxation time 
$\tau$, and it depends on the material properties, however, in this study the 
relaxation time is fixed to 10 fs ($\tau$ = 10$^{−14}$s) as taken by many 
studies.\cite{cem,ding} As seen in Fig.~\ref{5}, we calculated $S$, 
$\sigma$ and $S^2 \sigma$ for various temperature from 250K to 500K for every 
50K steps
using the BoltzTraP2 code. 
It should be noted, we only give the maximum value of the Seebeck coefficient and 
corresponding ($\sigma$) value for the selected temperatures. For bare XBi 
single-layers, we obtained a high Seebeck coefficient and the calculated largest 
$S$ value is for SnBi single-layer for all considered temperatures while the 
smallest $S$ value is for SiBi single-layer. With the increasing of the temperature 
from 250K to 500K the $S$ values of the XBi single-layers decrease more than half 
of the initial values. Electrical conductivity of XBi single-layers start to 
increase after 400K and the $\sigma$ of SiBi suddenly get a high value after 
450K. The efficiency of thermoelectric materials is given by the dimensionless 
figure of merit $ZT = S^2 \sigma T / \kappa$, where T is the absolute
temperature and $\kappa$ is the thermal conductivity, which is the sum of 
contributions from electron ($\kappa_e$) and lattice ($\kappa_l$) parts. High 
$ZT$ can obtain by a high power factor and a low $\kappa$ value. For this 
purpose we calculated the power factor of the XBi structures. As can be seen in 
the right part of the Fig.~\ref{5}, curve of the $S^2 \sigma$ is similar 
to the $\sigma$ curve due to dominant increasing of the $\sigma$. We note that 
the calculated $\kappa_e$ values are very low as needed for high $ZT$. Notably, in order to find an exact value for the $ZT$, the lattice thermal 
conductivity should be determined by the behavior of phonon transport in the XBi 
single-layers. Moreover, in terms of the electronic features, in the case of dopings, either $p-$ or $n-$type, semiconducting nature of 
single-layer SiBi will be affected in terms of the Fermi energy level. It can
be expected that the doping amount either increase or decrease the
thermoelectric power up to certain value of doping amount. Therefore, the effect doping on the thermoelectric properties can also be taken into account. 

Scattering region between the acoustic and optical phonon branches 
is important to determine the lattice thermal conductivity, large scattering 
region will lower the heat flux, and thus result in lower contribution of the 
thermal conductivity from phonons. 
As can be seen from the phonon structures of the XBi single-layers (see Fig.~\ref{1}(b)), 
frequency of the optical modes reduce while going from SiBi to SnBi and lowest optical branches 
of the XBi structures get closer to the highest acoustic branches which results in the narrower scattering region. 
The largest scattering region of the SnBi may give the lowest $\kappa_l$ value and it has relatively high $ZT$ value than other XBi single-layers.

\subsection{Substrate Effect}

\begin{figure}[!htb]
	\includegraphics[scale=1]{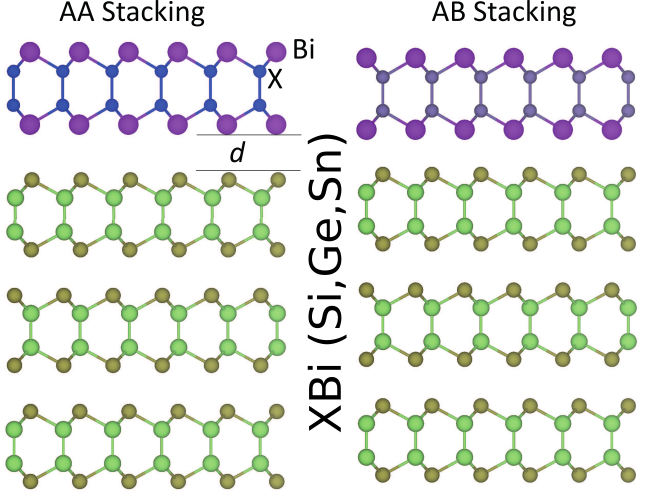}
	\caption{Side views of the XBi single-layers on various bulk structures. Two stacking types (AA stacking and AB stacking) considered according to top layer of the considered bulk materials.}	
	\label{6}
\end{figure}

As it is known, in order to synthesize a single-layer structure, a suitable substrate is of importance. There are two most important
restrictions which should be considered. First one is the lattice mismatch ratio between the substrate and the material to be 
synthesized, second is the interaction between the two materials that should be very 
weak as vdW binding to avoid the structural reconstructions on the single-layer structure. By considering these two condition we try to find suitable 
substrates for XBi single-layers. For this purpose we focused on the bulk materials 
which have similar atomic structure with XBi materials. The calculated lattice 
parameters are a=b=4.09 \AA{}\ and c=16.67 \AA{}\ for bulk GaTe, a=b=4.01 \AA{}\ 
and c=16.92 \AA{}\ for bulk InSe and a=b=4.25 \AA{}\ and c=17.69 \AA{}\ for bulk 
InTe. If we cut the bulk materials from the (001) plane, the lattice mismatch 
ratio between the unitcells of the bulk and XBi materials becomes less than 
2$\%$ which is acceptable ratio to growth XBi on these selected bulk materials. 
According to the lattice mismatch, GeBi placed on (001)GaTe, SiBi placed on both 
of (001)GaTe and (001)InSe and SnBi placed on (001)InTe substrates. Each XBi 
single-layers placed initially 3\AA{}\ above on the selected substrates and to 
avoid the computational cost three layers are selected for the substrates and 
all atoms released to geometric optimization. To determine the favorable 
stacking orientation we considered two stacking types as illustrated in 
Fig.~\ref{6}. The optimized structures revealed that AB stacking type is energetically 
approximately 0.20 eV more favorable than AA stacking order for the all structures. 
The calculated inter-layer energy between XBi and bulk 
materials for AB stacking types are -0.400 eV for GeBi@GaTe, -0.351 eV for 
SiBi@GaTe and -0.289 eV for SiBi@InSe and -0.409 for SnBi@InTe. These 
inter-layer energies indicate that the interaction
 between XBi and substrates can consider as vdW bonding, these results are 
verified by the Bader charge analysis, calculated charge transfer between the 
layers are in the range of 0.04-0.07 e$^-$. At the end of the optimization the 
normal distance in z-axis (\textit{d}) between XBi and substrates are in the range of 
3.10-3.40 \AA{}\. These results show that obtained properties for XBi single-layers 
will not effect much from these selected substrates due to physisorption of XBi 
single-layers and almost no net charge transfer between the XBi and substrates. 
In addition, the single-layer on top of the substrates are relaxed at 400 K for 
4 ps with \textit{ab-initio} molecular dynamics simulations and it is shown that 
the single-layers do not go under structural deformations. So, 
based on these results we can conclude that our predictive results of 
free-standing XBi outlined in the previous sections can be useful for 
experimentalists who attempts to synthesize these materials.

\section{Conclusion}

In this study, we predicted the dynamical stability of single-layer XBi (X=Ge, 
Si, or Sn) structures by means of first-principles calculations. Dynamical 
stability of each single-layer was verified in terms of their phonon band 
dispersions while the \textit{ab-initio} molecular dynamics simulations revealed the 
thermal stability of each single-layer structure. Raman spectra calculations 
indicated that three structures exhibit similarities in terms of their Raman 
active phonon modes with distinctive phonon frequencies. Electronic band 
structures showed that single-layer SiBi is an indirect band gap semiconductor 
while GeBi and SnBi exhibit metallic behaviors. Moreover, the calculated 
linear-elastic parameters indicated the quite soft nature of the proposed 
single-layers which makes them suitable for nanoelastic applications. Our 
predictions for the thermoelectric properties of single-layer XBi revealed the 
high potential of SiBi in terms of its thermoelectric coefficient which 
increases at high temperatures. Furthermore, we investigated the possible 
substrate effect on the structural formation of XBi single-layers. It was shown 
that either with their small lattice mismatch or weak substrate-XBi interaction, 
layered GaTe, InSe, and InTe are shown to be potential substrates for 
experimental realizations of single-layer XBi structures. Overall, it was shown 
that single-layer XBi structures can be alternative, stable 2D single-layers 
with their varying electronic and thermoelectric properties.

\section{Conflicts of interest}
The authors declare that there are no conflicts of interest regarding the 
publication of this paper.

\section{ACKNOWLEDGMENTS}
This work was supported by the National Research Foundation of Korea(NRF) grant funded by the Korea government(MSIT)(NRF-2017R1A2B2011989). 
Computational resources were provided by the Flemish Supercomputer Center (VSC). 
M.Y. is supported by the Flemish Science Foundation (FWO-Vl) by a postdoctoral 
fellowship.

\bibliography{ref}

\end{document}